\documentclass{ws-ijmpe}
\begin{document}
\markboth{B.K.Srivastava, R.P.Scharenberg and T.J.Tarnowsky}{Understanding the Particle Production Mechanism with Correlation Studies Using Long and Short Range Correlations
}

\catchline{}{}{}{}{}
%%%%%%%%%%%%%%%%%%%%%%%%%%%%%%%%%%%%%%%%%%%%%%%%%%%%%%%%%%%%%%%%%%%%

\title{UNDERSTANDING THE PARTICLE PRODUCTION MECHANISM WITH CORRELATION STUDIES USING LONG AND SHORT RANGE CORRELATIONS}

\author{{B. K. SRIVASTAVA, R. P. SCHARENBERG AND  T. J. TARNOWSKY\\(for the STAR Collaboration)}}

\address{Department of Physics, Purdue University, \\
West Lafayette, Indiana-47907,
USA\\
brijesh@physics.purdue.edu}

\maketitle

\begin{history}
\received{(received date)}
\revised{(revised date)}
%\accepted{(Day Month Year)}
%\comby{(xxxxxxxxxx)}
\end{history}

\begin{abstract}
Long range forward-backward multiplicity correlations have been measured with the STAR detector for  Au+Au collisions at $\sqrt{s_{NN}}$ = 200 GeV . Strong long range correlations are observed in central Au+Au collisions. Based on the Dual Parton model and Color Glass Condensate considerations the data suggests that these long range correlations are due to multiple parton interactions. This suggests that dense partonic matter is created in  central Au+Au collisions at $\sqrt{s_{NN}}$ = 200 GeV.

\end{abstract}

\section{Introduction}
The study of correlations among particles produced in different rapidity
regions may provide understanding of the mechanisms of particle production. 
Strong short-range correlations (SRC) are observed in hardon-hadron collisions, 
indicating clustering over a region of $\sim \pm$1 unit in rapidity \cite{uhling,alner,ansorge,aexopoulos}. 
Correlations that extend over a longer range are observed in hadron-hadron
interactions only at higher energies \cite{alner,aexopoulos}. 

Multiparticle production can be described in terms of color strings stretched between 
the oncoming partons. These strings then decay into observed secondary hadrons. In this 
scenario, the strings emit particles independently \cite{capela1,capela2,kaidalov}. 
It has been suggested that long-range correlations (LRC) might be enhanced
in hadron-nucleus and nucleus-nucleus interactions, compared to hadron-hadron scattering at the same energy \cite{capela1,capela2}.   
One way to study the correlation is to measure forward-backward multiplicity correlations which have been studied in several experiments \cite{uhling,alner,ansorge,aexopoulos}.

The correlation strength is defined by the dependence of the average charged particle multiplicity in the backward hemisphere $\langle N_{b}\rangle$, on the
event multiplicity in the forward hemisphere $N_{f}$, $\langle N_{b}\rangle$=a+$b$$N_{f}$, where a is a constant and $b$ measures the strength of the correlation\cite{capela1,capela2}:  
\begin{equation}
 b = \frac{\langle N_{f}N_{b}\rangle - \langle N_{f}\rangle \langle N_{b}\rangle}{\langle N_{f}^{2}\rangle - \langle N_{f} \rangle ^{2}}= \frac{D_{bf}^{2}}{D_{ff}^{2}} 
\label{b}
\end{equation}  
In ``Eq.(1)'' $D_{bf}^{2}$ and $D_{ff}^{2}$ are the backward-forward and forward-forward dispersions respectively.

\section{Data Analysis}

The data utilized for this analysis is  for  Au+Au collisions at $\sqrt{s_{NN}}$ = 200 GeV at the Relativistic Heavy Ion Collider (RHIC), as measured by the STAR experiment \cite{starnim}. The collision events were part of the minimum bias data set. The centralities presented in this analysis account for 0-10\%, 10-20\%, 20-30\%, 30-40\% and 40-50\% of the total hadronic cross section and were obtained by use of 
the multiplicity of all charged particles ($N_{ch}$) measured in the TPC with $|\eta| < $0.5. 
 All charged particles in the Time Projection Chamber (TPC) pseudorapidity range $|\eta| < 1.0$ and with $p_{T} >$ 0.15 GeV/c were considered. This region was subdivided into bin widths $\eta = 0.2$. The forward-backward intervals were located symmetrically about midrapidity with the distance between bin centers ($\Delta\eta$): 0.2, 0.4, 0.6, 0.8, 1.0, 1.2, 1.4, 1.6, and 1.8. 
An analysis of the data from \textit{pp} collisions at 200 GeV was also performed on minimum bias events using the same quality cuts as in the case of Au+Au.
Corrections for detector geometric acceptance and tracking efficiency were carried out using a Monte Carlo event generator and propagating the simulated particles through a GEANT representation of the STAR detector geometry. 
\begin{figure}[th]
\centerline{\psfig{file=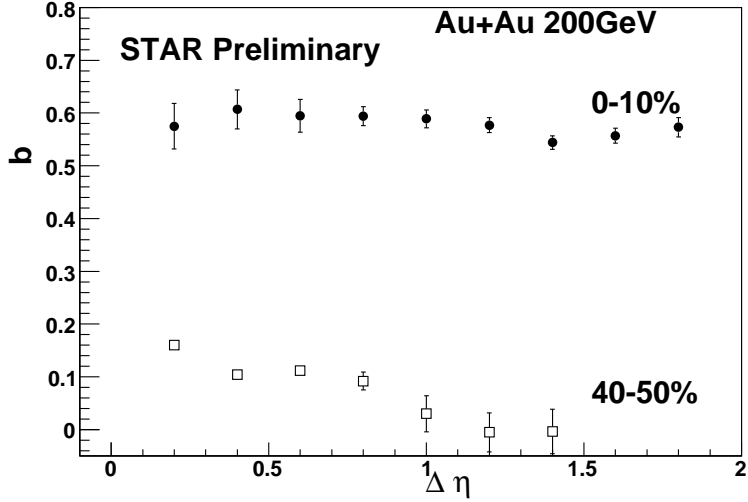,width=11cm}}
\vspace*{8pt}
\caption{ Correlation strength $b$ for 0-10\% and 40-50\% centrality  as a function of pseudorapidity gap $\Delta\eta$ for Au+Au collisions at $\sqrt{s_{NN}}$ = 200 GeV.}
\label{CentralAu}
\end{figure}

In order to eliminate (or at least reduce) the effect of impact parameter (centrality) fluctuations on this measurement, each relevant quantity ($N_{f}$, $N_{b}$, $N_{f}^{2}$,  $N_{f}N_{b}$) was obtained on an event-by-event basis as a function of $N_{ch}$, and was fitted to obtain the average values of
$\left<N_{f}\right>$,
$\left<N_{b}\right>$, $\left<N_{f}\right>^{2}$, and $\left<N_{f}N_{b}\right>$ \cite{ebye1,ebye2}. This method removes the dependence of the correlation strength b, on the size of the centrality bin $N_{ch}$. 
Tracking efficiency and acceptance corrections were applied to each event. 
Systematic effects dominate the error determination. The systematic errors are determined by varying cuts on the z-vertex $|v_{z}|$ ($<$ 30, 20, and 10 cm), number of fit points on the individual tracks in the TPC.

\section{Results and Discussions}

The plot of correlation strength ($b$) as a function of the pseudorapidity gap is shown in Fig. \ref{CentralAu} for the 0-10\% most central Au+Au events. It is observed that the value of $b$ does not change with the pseudorapidity gap.  Fig. 1 also shows $b$ as a function of $\Delta\eta$ for 40-50\% mid-central Au+Au collisions. In this case $b$ decreases with the increasing $\Delta\eta$, which is expected if there were only short range correlations.

\begin{figure}[th]
\centerline{\psfig{file=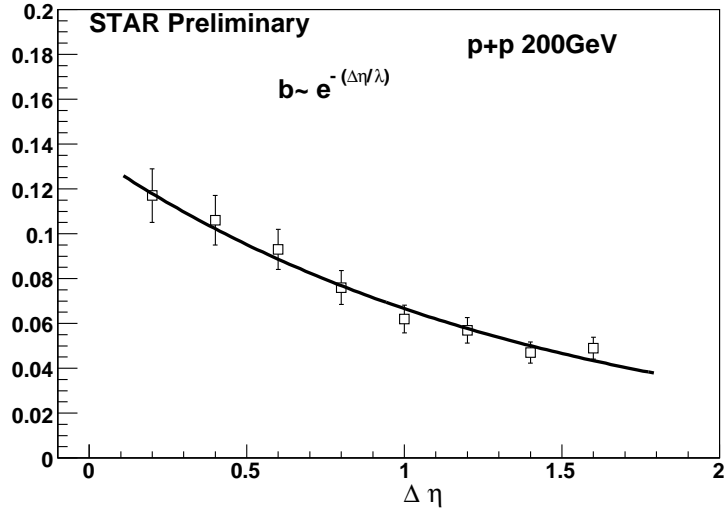,width=11cm}}
\vspace*{8pt}    
\caption{(a) Correlation strength b, and the line is the exponential fit to b vs $\Delta\eta$.}
\label{PeripAuandpp}
\end{figure}

Short range correlations have been extensively studied. The shape of the SRC function has a maximum at, and is symmetric about, midrapidity (pseudorapidity, $\eta$ = 0). It can be fitted with a Gaussian or an exponential function \cite{ansorge,capela1}.
 The short range correlations  are significantly reduced 
by a separation of 1.5 - 2.0 units of pseudorapidity \cite{capela2}.
  Fig. \ref{CentralAu} shows that in case of 0-10\% most central Au+Au the $b$ value is flat with $\Delta\eta$, while the mid central 40-50\% has a sharp fall like SRC.  Since the analysis is limited to $-1<\eta<1$, the short range component cannot be completely eliminated. The \textit{pp} collisions is used to estimate the short range component from Au+Au collisions. Fig. \ref{PeripAuandpp} shows $b$ as a function of $\Delta\eta$ for the \textit{pp} collisions. The $b$ value from \textit{pp} decays exponentially with increasing $\Delta\eta$.  Fig. \ref{CentralAu} and Fig. \ref{PeripAuandpp} show that the correlation strength with $\Delta\eta$ is quite different in 0-10\% Au+Au as compared to \textit{pp}.  
An analysis of the mid-central and peripheral events in Au+Au shows that for the 40-50\% centrality bin the shape of correlation strength b with $\Delta\eta$ is similar to the \textit{pp} case.

\begin{figure}[th]
\centerline{\psfig{file=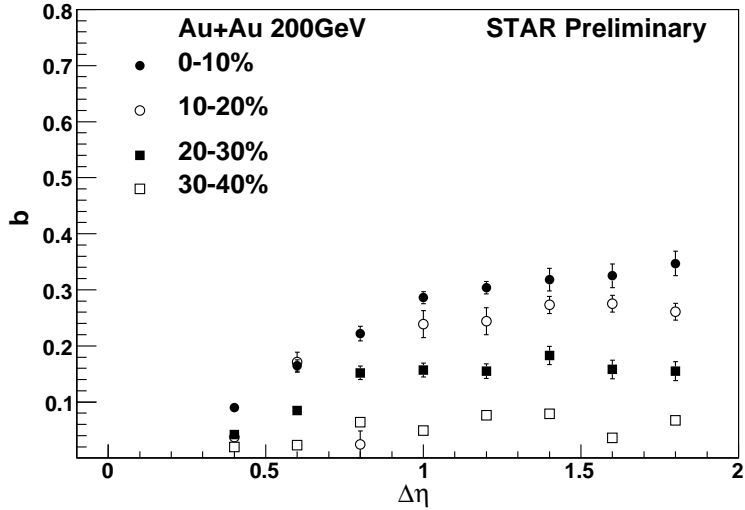,width=11cm}}
\vspace*{8pt}    
\caption{ Growth of LRC  for centrality bins 0-10\%, 10-20\%, 20-30\% and 30-40\%.}
\label{CentEvolution}
\end{figure}

Traditionally, the LRC is measured with a large gap in the forward-backward window and is based on  
experimental observation of charged particle correlations in $p \bar{p}$ at $\sqrt {s_{NN}}=$  200, 500, and 900 GeV \cite{ansorge}. The strength of the SRC is reduced considerably with a $\Delta\eta$ of 2.0 units. Thus the remaining portion of the correlation strength is mainly due to the LRC. 
As shown in Fig. \ref{PeripAuandpp}, it appears that \textit{pp} at 200 GeV only has a SRC.
We assume that the fall of  SRC as a function of $\Delta\eta$ does not change in going from $\it pp$ to Au+Au. The 40-50\% Au+Au supports this assumption.
To extract the short range part in 0-10\% Au+Au collisions, the \textit{pp} value of $b$ at $\Delta\eta=0$ (non-overlapping measurement windows), is scaled up to the $b$ value in the Au+Au case, as the SRC has a maximum at $\eta=0$. $b$ vs $\Delta\eta$  in Fig. \ref{PeripAuandpp} for \textit{pp} was fitted with an exponential function. The scaled $b$ value for the other $\Delta\eta$ points were calculated using the fit function.
The remaining part of the correlation strength (long range), is obtained by subtracting  the short range component from the measured correlation strength. This is shown in Fig. \ref{CentEvolution} as the long range. There is a growth of the LRC with increasing $\Delta\eta$. 

Two other analyses in STAR have focused on the correlations of charged particle pairs in $\Delta\eta$ ( pseudorapidity) and $\Delta\phi$ (azimuth) \cite{tom,ron}. It was observed that near-side peak is elongated in $\Delta\eta$ in central Au+Au as compared to peripheral collisions.

\begin{figure}[th]
\centerline{\psfig{file=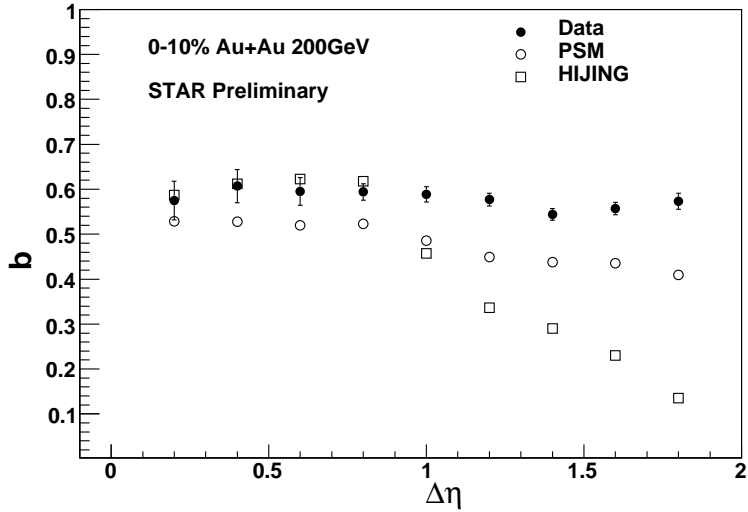,width=11cm}}
\vspace*{8pt}    
\caption{Model comparison with data. The correlation strength is shown for HIJING (open square) and the Parton String Model, PSM(open circle) for the 0-10\% centrality in Au+Au collisions.}
\label{Models}
\end{figure}

  The centrality of the collision plays an important role in the growth of long range component of the total correlation strength. Data from 10-20\%, 20-30\%, and 30-40\% most central Au+Au collisions have been analyzed, following the same procedure as for 0-10\% centrality, to determine the evolution of the LRC strength. The growth of LRC is shown in Fig \ref{CentEvolution}. The magnitude of the LRC is quite large for the most central collisions when $ \Delta\eta >$ 1.0. From Fig. \ref{CentEvolution} it is clear that the magnitude of the LRC increases from peripheral to central collisions. 
      The 0-10\% results are also compared with phenomenological models HIJING \cite{hijing} and the DPM \cite{capela2}. Monte Carlo codes HIJING and the Parton String Model (PSM) \cite{capela2,amelin2,armesto3} were used to generate minimum bias events for Au+Au collisions at 200 GeV. The PSM is based on DPM \cite{capela2}. The analysis was carried out in the same manner as with real data. The PSM was used without the string fusion option. The variation of b with $\Delta\eta$ is shown in Fig. \ref{Models} along with the experimental  value for 0-10\% central Au+Au collisions. 

HIJING predicts SRC with a large value of $b$ near midrapidity in agreement with the data. A sharp decrease in $b$ is seen beyond the $\Delta\eta \sim$ 1.0. PSM has both short and long range correlations and is in qualitative agreement with the data. 

In the DPM the particle production occurs via string fragmentation. There are two strings per inelastic collision and the long range component is  expressed as
\begin{equation}
\langle N_{f}N_{b}\rangle - \langle N_{f}\rangle \langle N_{b}\rangle \propto [\langle n^{2}\rangle - {\langle n \rangle}^{2}] {\langle N_{q-\overline{q}}\rangle}_{f}  {\langle N_{q-\overline{q}}\rangle}_{b}
\label{LRC} 
\end{equation}   
where the average multiplicities of $q-\overline{q}$ in the forward and backward regions is given by $ {\langle N_{q-\overline{q}}\rangle}_{f}$ and  $ {\langle N_{q-\overline{q}}\rangle}_{b}$ respectively in each elementary inelastic collision. ``Eq. (\ref{LRC})'' shows that the LRC is due to fluctuations in the number of elementary inelastic collisions. It is believed that the experimental observation of the LRC originates from these multiple partonic interactions. 
The Color Glass Condensate (CGC) provides a theoretical QCD based description of multiple string interactions \cite{venu,iancu}. The CGC \cite{larry} argues for the existence of a LRC in rapidity, similar to those predicted in DPM.  Recently, long range forward-backward multiplicity correlations have been discussed in the framework of the CGC and predict the growth of the LRC with the centrality of the collision \cite{larry2}.  

\section{Summary}
In summary, this is the first work on the measurement of the long-range correlation strength ($b$), in ultra relativistic nucleus-nucleus collisions.   
The DPM and CGC argue that the long range correlations are due to multiple parton-parton interactions. This indicates that dense quark-gluon matter is formed in central Au+Au collisions at $\sqrt{s_{NN}}$ = 200 GeV. An analysis of the correlation for baryons can further address the CGC picture.

\end{document}